\documentclass[11pt,a4paper]{article}
\pdfoutput=1

\usepackage{jcappub}
\usepackage[utf8]{inputenc}
\usepackage{multirow}
\usepackage{hyperref}

\newcommand{\ECR}{E_{\rm CR}}

\title{Cosmogenic Neutrinos Through the GRAND Lens Unveil the Nature of Cosmic Accelerators}
\author[a]{Klaes M\o ller,}
\emailAdd{xgb609@alumni.ku.dk}
\author[a]{Peter B.~Denton}
\emailAdd{denton@nbi.ku.dk}
\author[a]{and Irene Tamborra}
\emailAdd{tamborra@nbi.ku.dk}

\affiliation[a]{Niels Bohr International Academy and DARK, Niels Bohr Institute, University of Copenhagen, Blegdamsvej 17, 2100, Copenhagen, Denmark}

\abstract{The sources of cosmic rays with energies above 55 EeV are still mysterious.
A guaranteed associated flux of ultra high energy neutrinos known as the cosmogenic neutrino flux will be measured by next generation radio facilities, such as the proposed Giant Radio Array for Neutrino Detection (GRAND).
By using the orthogonal information provided by the cosmogenic neutrino flux, we here determine the prospects of GRAND to constrain the source redshift evolution and the chemical composition of the cosmic ray sources.
If the redshift evolution is known, independently on GRAND's energy resolution, GRAND with 200,000 antennas will constrain the proton/iron fraction to the $\sim5-10\%$ level after one year of data taking; on the other hand, if hints on the average source composition are given, GRAND will measure the redshift evolution of the sources to a $\sim 10\%$ uncertainty.
However, the foreseen configuration of GRAND alone will not be able to break the degeneracy between redshift evolution of the sources and their composition. Our findings underline the discriminating potential of next generation radio array detectors and motivate further efforts in this direction.}

\begin{document}

\maketitle

\section{Introduction}
The cosmic ray spectrum has been measured up to energies of $\ECR>100$ EeV (1 EeV = $10^9$ GeV) by Pierre Auger Observatory \cite{Fenu:2017hlc} and Telescope Array (TA) \cite{Ikeda:2017dci}.
Yet, the nature of the sources of extreme energy cosmic rays (EECRs\footnote{The Ultra High Energy Cosmic Ray (UHECR) typically refers to cosmic rays with energies $\gtrsim1$ EeV. We use the term EECR adopted by the community to mean those with energies $\gtrsim55$ EeV which have a short horizon.}) has been puzzling astronomers for more than a century~\cite{Anchordoqui:2018qom}. 

Due to interactions with photons from the Cosmic Microwave Background (CMB) and Extragalactic Background Light (EBL), the Universe is opaque to cosmic rays with energies $\ECR\gtrsim55$ EeV on distance scales $\gtrsim100$ Mpc \cite{Greisen:1966jv,Zatsepin:1966jv} which should make source identification through anisotropy possible.
Despite the fact that the local Universe is fairly anisotropic, the sources of these EECRs remain hidden, likely due to bending in intervening magnetic fields \cite{Waxman:1996zn,Jansson:2012rt,Ahlers:2017wpb}.
In addition, the composition of the EECRs is uncertain \cite{Aab:2014kda,Aab:2017cgk} which, along with the uncertain galactic and extragalactic magnetic fields, adds a further complication to measuring the EECR anisotropy and discovering point sources.

One path to understanding the nature of the EECR sources is by comparing the measured energy spectrum of the cosmic rays with theoretical predictions.
The energy spectrum, however, is sharply modified by interactions with photons which mask the initial spectrum and lead to a strong suppression of the EECR flux at $\ECR\gtrsim55$ EeV.

Three main alternative ways have been considered to probe the composition of EECRs.
The first involves measuring the properties of air showers when cosmic rays interact in the atmosphere, and has been used extensively by Auger and TA.
This method relies on the measurement of the depth into the atmosphere at which the maximum fluorescence light is produced, $X_{\max}$, and comparing this to simulations using hadronic models, which still have considerable uncertainties \cite{Pierog:2018nkf,Aab:2017cgk}.
Additionally, the muon flux from the shower can also be used to determine the evolution of the shower through $X_{\max}^\mu$, but there the hadronic models generally do not describe the data well.
These air shower measurements tend to prefer a light composition that increases in mass with the energy of the cosmic rays, although with considerable uncertainties.
The next option is to compare the measured EECR isotropy with the known anisotropy of the local Universe.
In this case one sees that a light composition is disfavored in order to produce the low levels of anisotropy observed at the Earth \cite{Ahlers:2017wpb}.
The final option is the one this paper focuses on: to measure the neutrinos produced as byproducts from EECRs scattering off the CMB and EBL photons which is known as the cosmogenic neutrino flux, and is related to the cosmic ray flux from larger distances than direct EECR measurements.

Detecting the cosmogenic flux of neutrinos requires massive detectors beyond what is currently available. In fact, current generation high energy neutrino experiments \cite{Aartsen:2018vtx,Aartsen:2016ngq,Allison:2018cxu,Aab:2015kma,Allison:2015eky,Nelles:2016fxe} are not likely to detect this flux except in the most optimistic of scenarios.
Future experiments such as the Giant Radio Array for Neutrino Detection (GRAND), which is currently designed to cover an area of 200,000 km$^2$ with 200,000 radio antennas \cite{Fang:2017mhl}, and the Probe Of Extreme Multi-Messenger Astrophysics (POEMMA) with a comparable effective area via two satellites measuring fluorescence light \cite{Olinto:2017xbi}, should measure the cosmogenic neutrino flux independently of the source model parameters~\cite{AlvesBatista:2018zui}.
Two current generation radio experiments are ARA \cite{Allison:2015eky} and ARIANNA \cite{Nelles:2016fxe}.
ARA leverages the Askaryan effect to identify ultra high energy neutrinos in the ice at the South Pole and ARIANNA has a similar program on the Ross Ice Shelf.
In this article we will focus on GRAND to provide a concrete example, but our results will generally apply to any upcoming large-scale ultra high energy neutrino experiment.

Predictions of the expected cosmogenic neutrino flux have been presented in the literature already, see e.g.~\cite{Ahlers:2012rz,Aloisio:2015ega,AlvesBatista:2018zui}. In this article, we fit the spectrum normalization and its injection spectral index to Auger's EECR data and employ the cosmogenic neutrino flux to extrapolate information on the redshift evolution of the EECRs sources as well as the EECR composition.
Then we look at how this feeds back into our global understanding of cosmic ray sources using both EECR and cosmogenic neutrinos.

The layout of this article is as follows. In Sec.~\ref{sec:interplay}, the mechanisms behind the neutrino production as a function of the EECR source composition are outlined. General details on the detection perspectives of the cosmogenic neutrino flux in the currently planned GRAND layout with 200,000 radio antennas GRAND will be discussed in Sec.~\ref{sec:detection}. Section~\ref{sec:analysis} focuses on what can be learned about EECRs and their sources from a measurement of the cosmogenic neutrino flux with GRAND. Sections \ref{sec:results} and \ref{sec:discussion} discuss our findings, while conclusions and perspectives are reported in Sec.~\ref{sec:conclusions}. Finally, intermediate EECR source composition scenarios between the extreme cases considered in the paper (EECRs purely made by protons or heavy nuclei) are discussed in Appendix~\ref{sec:intermediate composition}.

\section{Interplay of Extreme Energy Cosmic Rays and Cosmogenic Neutrinos}
\label{sec:interplay}

The rate of neutrino production varies with the chemical composition of the EECRs en route from the accelerator to the Earth. Although any possible source composition is allowed, for the sake of simplicity, we here distinguish two separate classes: sources mainly composed by protons and sources mainly containing heavy nuclei. The correspondent EECR and neutrino production are described below as well as the neutrino flavor ratio expected on Earth.

Due to photopion production and photodisintegration of heavy nuclei along with pair production, the maximum distance that EECRs travel is limited meaning that they are a probe of the local Universe out to $z\sim0.02$ ($\sim100$ Mpc).
Neutrinos, on the other hand, pass through the Universe relatively undisturbed\footnote{At $E_\nu\gtrsim10^4$ EeV neutrinos resonantly scatter off the C$\nu$B through an on-shell Z boson \cite{Weiler:1982qy}. These energies are well beyond the peak of the cosmogenic neutrino flux.} and only suffer energy loss due to cosmological expansion.
As such, neutrinos provide a separate lever arm to probe the redshift evolution of cosmic accelerators.

\subsection{Source composition}\label{sec:composition}
If the sources of EECRs are mainly protons, at around $E_p\simeq55$ EeV, protons scattering off CMB photons have enough energy in the center-of-mass frame to resonantly produce a $\Delta^+$ baryon which leads to a considerably increased cross section.
The $\Delta^+$ has enough mass ($m_{\Delta^+}=1232$ MeV) to dominantly decay to a pion and a nucleon with two-thirds of the decays to $\pi^0+p$ and one-third of the decays to $\pi^++n$ where the branching ratios are given by Clebsch-Gordan coefficients.
The neutral pion decays dominantly to a pair of photons which will cascade down and could potentially be constrained by {\it Fermi}-LAT \cite{Ackermann:2014usa} measurements of the extragalactic gamma-ray flux \cite{Hooper:2010ze,Berezinsky:2016jys,Supanitsky:2016gke,vanVliet:2016dyx,Globus:2017ehu,vanVliet:2017obm}.
The charged pion decays to $\nu_e$, $\nu_\mu$, $\bar\nu_\mu$, and $e^+$ providing a guaranteed flux of ultra high energy neutrinos.
The resultant nucleon carries away only $\sim80\%$ of the energy which leads to the notion of the Greisen-Zatsepin-Kuzmin (GZK) horizon which limits the distance that EECRs can travel in the CMB photon field to be $\sim100$ Mpc \cite{Greisen:1966jv,Zatsepin:1966jv}.

Another possibility is that the EECR sources would be mainly constituted of heavy nuclei.
In that case, the photon-nucleus cross section is dominated by the giant dipole resonance for photons with center-of-mass energies 8 MeV $<E_\gamma^\prime<$ 30 MeV.
The giant dipole resonance contributes the largest to the cross section since the photons interact with the entire nucleus which behaves as a fluid \cite{Puget:1976nz,Aloisio:2017qoo}.
This process leads to the loss of one or more nucleons from the nucleus and then subsequent decays of the likely unstable daughter nucleus, as well as the neutrons, all of which lead to a flux of ultra high energy neutrinos. 
In addition, the resultant protons also undergo photopion production.
Due to the energetics of these processes, the neutrino flux from heavy nuclei tend to be less than that from lighter nuclei.
When the EECR spectrum is fixed, the neutrino flux from protons is comparable to that of an intermediate mass nucleus, see Appendix \ref{sec:intermediate composition} and Fig.~\ref{fig:intermediate composition}.

\subsection{Neutrino flavor composition}
\label{ssec:flavor}
An important indicator of the source physics is the neutrino flavor ratio. According to the physics described above, the benchmark assumption is that the flavor ratio of ultra high energy astrophysical neutrinos is $(N_{\nu_e}:N_{\nu_\mu}:N_{\nu_\tau})_\oplus=1:1:1$ at the Earth\footnote{Here we are referring to the sum of neutrinos and anti-neutrinos. Distinguishing between the two at $E_\nu\gtrsim10$ PeV is essentially impossible.}.
In fact, to first order, at the source (S) the reactions introduced in Sec.~\ref{sec:composition} produce neutrinos with an initial flavor ratio of $(N_{\nu_e}:N_{\nu_\mu}:N_{\nu_\tau})_S= 1:2:0$ from charged pion decay which oscillates to 1:1:1 after long distances under the tri-bimaximal flavor structure; this is a good approximation for distance averaged oscillations \cite{Farzan:2008eg,Anchordoqui:2013dnh}.

\begin{figure}
\centering
\includegraphics[width=.9\textwidth]{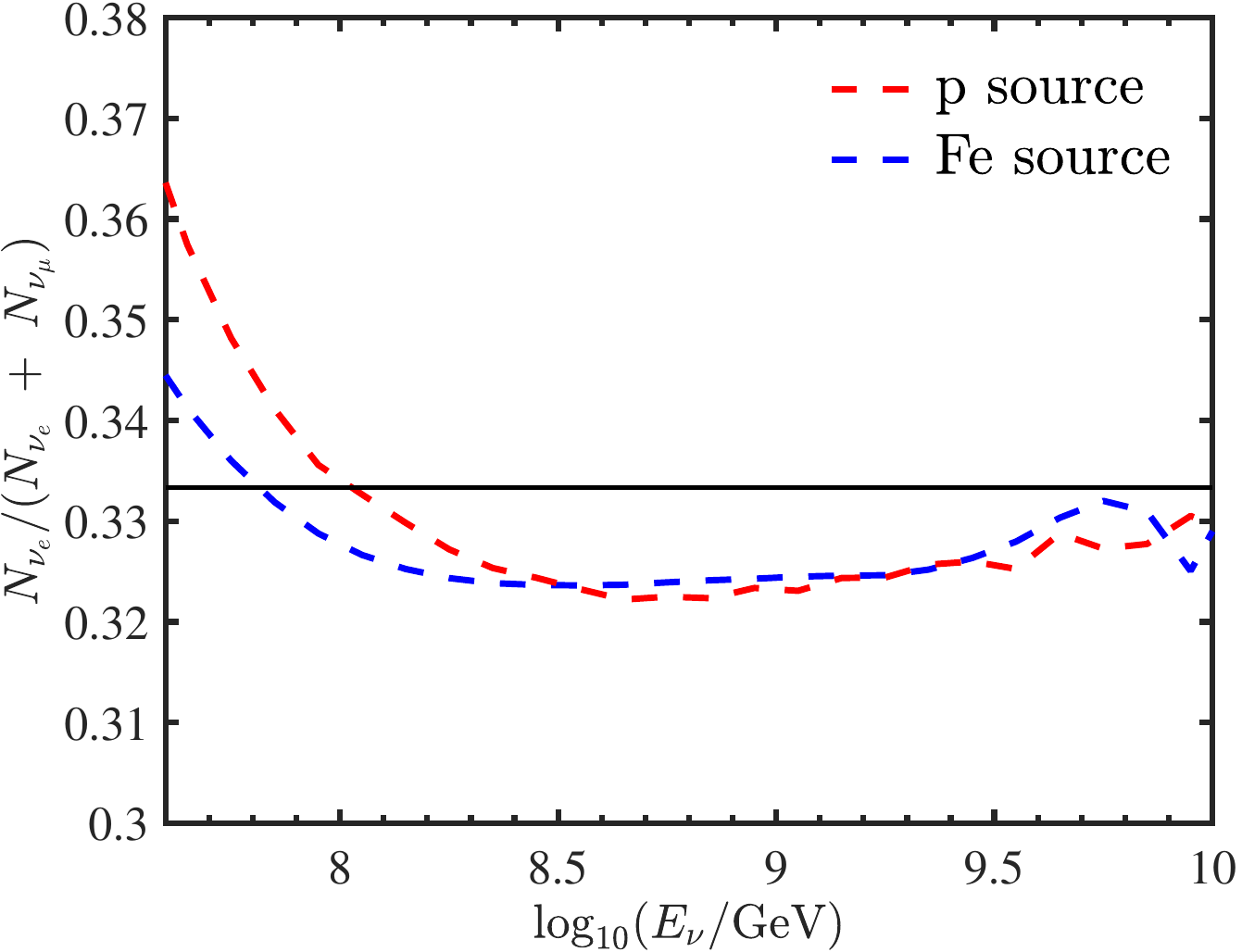}
\caption{Flavor ratio at production $N_{\nu_e}/(N_{\nu_e}+N_{\nu_\mu})$ for an EECR flux mainly composed by protons (in red) or iron (in blue), as a function of the neutrino energy.
The flavor ratio is nearly 1/3 (consistent with 1:2:0) as expected.
The slight deficit of neutrinos in the central energy bins is from neutrons that are then regenerated at low energies, below the reach of GRAND.
The minimum flavor ratio is $0.32$ which justifies the usage of 1:1:1 flavor ratio at the Earth; the latter is represented trough the black horizontal line to guide the eye.}
\label{fig:flavor}
\end{figure}

Certain processes may alter this flavor ratio.
The most prominent correction is from neutron decay which occurs since neutrons are produced as a part of photopion production and photodisintegration.
This leads to an additional flux of $\bar\nu_e$, although the energy of the resultant $\bar\nu_e$ is suppressed by $\gtrsim2$ orders of magnitude compared to the neutrinos coming from charged pion decay.

In order to verify that the corrections to the conventional picture of neutrinos from $\pi$-decay are small, we perform a numerical computation.
Figure \ref{fig:flavor} shows the flavor ratio of cosmogenic neutrinos expected at source, in blue for an iron source and in red for a proton source as calculated with CRPropa3 \cite{Batista:2016yrx}.
The upturn at low energies arises from neutron decay; however, this would not be testable with GRAND which is only sensitive to $\nu_\tau$'s (see Sec.~\ref{sec:detection}). One can notice that the impact of the EECR source composition on the resultant flavor ratio is negligible. Moreover, the minimum expected flavor ratio is $0.32$ which justifies the usage of $N_{\nu_e}:N_{\nu_\mu}:N_{\nu_\tau} = 1:1:1$ flavor ratio at the Earth (plotted in black in Fig.~\ref{fig:flavor}).

\section{Detection of Ultra High Energy Neutrinos in GRAND}
\label{sec:detection}
Numerous projects aim to detect ultra high energy neutrinos including IceCube \cite{Aartsen:2018vtx}, ANITA \cite{Allison:2018cxu}, Pierre Auger Observatory \cite{Aab:2015kma}, ARA \cite{Allison:2015eky}, and ARIANNA \cite{Nelles:2016fxe} along with several future experiments, two of which are POEMMA \cite{Olinto:2017xbi} and GRAND \cite{Fang:2017mhl}.
Several different techniques can be employed to measure ultra high energy neutrinos; in this work, we will focus on the Earth-skimming $\nu_\tau$ technique.
This technique uses the fact that any extensive-air-shower that emerges from the Earth is best explained within the neutrino Standard Model as an ultra high energy up-going $\nu_\tau$ that interacts near the surface of the Earth creating a $\tau$ which escapes the Earth before decaying (note that neutrino oscillation effects in the Earth are negligible for $E_\nu\gtrsim1$ TeV).
The Earth-skimming signature can be used to identify ultra high energy neutrinos as neutrino absorption through the whole Earth at these energies is considerable.
To further improve their sensitivity, some experiments (Auger and GRAND) take advantage of mountainous terrain which still provides discrimination between showers from ultra high energy neutrinos and EECRs while minimizing the amount of material ultra high energy neutrinos must survive before interacting~\cite{Aab:2015kma,Fang:2017mhl}.

After the $\tau$ escapes the Earth, it decays creating an Earth-skimming extensive air shower.
As the shower propagates, coherent radio emission is produced.
GRAND plans to cover a very large area to detect this signal with 200,000 radio towers in mountainous regions to detect showers from $\tau$ decay emerging from mountains and has already demonstrated air shower detection as a proof-of-principle with the TREND array \cite{Charrier:2018fle}.

GRAND expects to have some energy resolution capabilities for the hadronic shower \cite{Fang:2017mhl}.
Reconstructing the energy of the initial neutrino, however, is more complicated and will require an unfolding process to account for the energy loss of the neutrino in matter, the fraction of energy transferred to the $\tau$, and the energy loss of the $\tau$ before escaping the Earth.
This process will also require an assumed initial neutrino spectrum.
We take the uncertainty on the neutrino energy to be $\Delta\log_{10}E_\nu=0.25$ which is equivalent to $^{+78\%}_{-44\%}$.
Given that GRAND should be able to reach $15\%$ shower energy resolution, once unfolding into the true neutrino energy is accounted for this energy estimate should be achievable \cite{GRAND_whitepaper}.

In this work, unless otherwise specified, we calculate the observed event rate at GRAND using the foreseen effective area with one year of 200,000 km$^2$ exposure~\cite{Martineau-Huynh:2015hae,GRAND_whitepaper}.
This assumes a zenith angle region of $85^\circ\le\theta_z\le95^\circ$ which covers $\sim80\%$ of the sky every day for the single site configuration (multiple sites could cover the entire sky depending on their location).

\section{Source Population Modeling}
\label{sec:analysis}
Our analysis focuses on what can be learned about EECRs and their sources from a measurement with GRAND of cosmogenic neutrinos.
Given the inconsistency of the EECR flux normalization and composition provided by current data \cite{TheTelescopeArray:2018dje}, in this work we use as little input from other experiments as possible to reduce the systematic uncertainties that appear in each detection channel.
Unless otherwise specified, the only non-GRAND information used is the EECR energy spectrum from Auger \cite{Fenu:2017hlc}.

Using CRPropa3 \cite{Batista:2016yrx} with the EBL model from Ref.~\cite{Finke:2009xi}, we calculate the expected flux of EECRs and ultra high energy neutrinos as a function of source parameters.
We account for energy loss from pair production, photopion production, photodisintegration, and redshift evolution, and we track all nuclei and neutrinos.

We assume that each source class follows a redshift evolution inspired by that of the star formation rate \cite{Yuksel:2008cu}
\begin{equation}
\rho(z) \propto
\begin{cases}
(1+z)^m&z<1\\
2^m&1<z<4\\
2^m(1+z)^{-3.5}\quad&z>4
\end{cases}\,,
\end{equation}
up to maximum redshift $z_{\max}=7$ where $m$ is a continuous parameter that we will leave free to vary in this work.
The source composition is expressed in terms of $\alpha_S$, in such a way that the total flux $F_S$ at the source is
\begin{equation}
F_S = (1-\alpha_S) F_p+\alpha_S F_{\rm Fe}\,,
\label{eq:alphaS}
\end{equation}
where $\alpha_S$ parameterizes the fraction of iron in the source ($\alpha_S = 0$ corresponding to a proton source). We also assume that the flux follows a power law $F_p \propto E_p^{-\gamma}$ (with $E_p$ being the proton energy) and similarly $F_{\rm Fe}$.
We continue the spectrum up to rigidity $250$ EV where we apply a hard cutoff.
We also assume that there is no modification of the spectrum due to diffusion in magnetic fields \cite{Mollerach:2013dza} since the effect seems to be small at these extreme energies \cite{Batista:2014xza}.
As we will see later, this choice of parameterization of the composition is actually conservative in our ability to estimate the sensitivity to the composition as the intensity of neutrinos from a flux of light nuclei is actually higher than that from protons, see Appendix \ref{sec:intermediate composition}.

Hence, the intensity of cosmogenic neutrinos is parametrized as a function of four source parameters: the slope of the redshift evolution $m$, the source composition $\alpha_S$, the source spectral index $\gamma$, and the flux normalization $\Phi_0$ which contains the source density and typical source luminosity. Note that, for fixed source class, $\alpha_S$ could be source-dependent in principle given the composition of the host environment. Here, we assume $\alpha_S$ to be the average fraction of iron representative of the whole source population and assume no redshift dependence on $\alpha_S$.
\begin{figure}
\centering
\includegraphics[width=.9\textwidth]{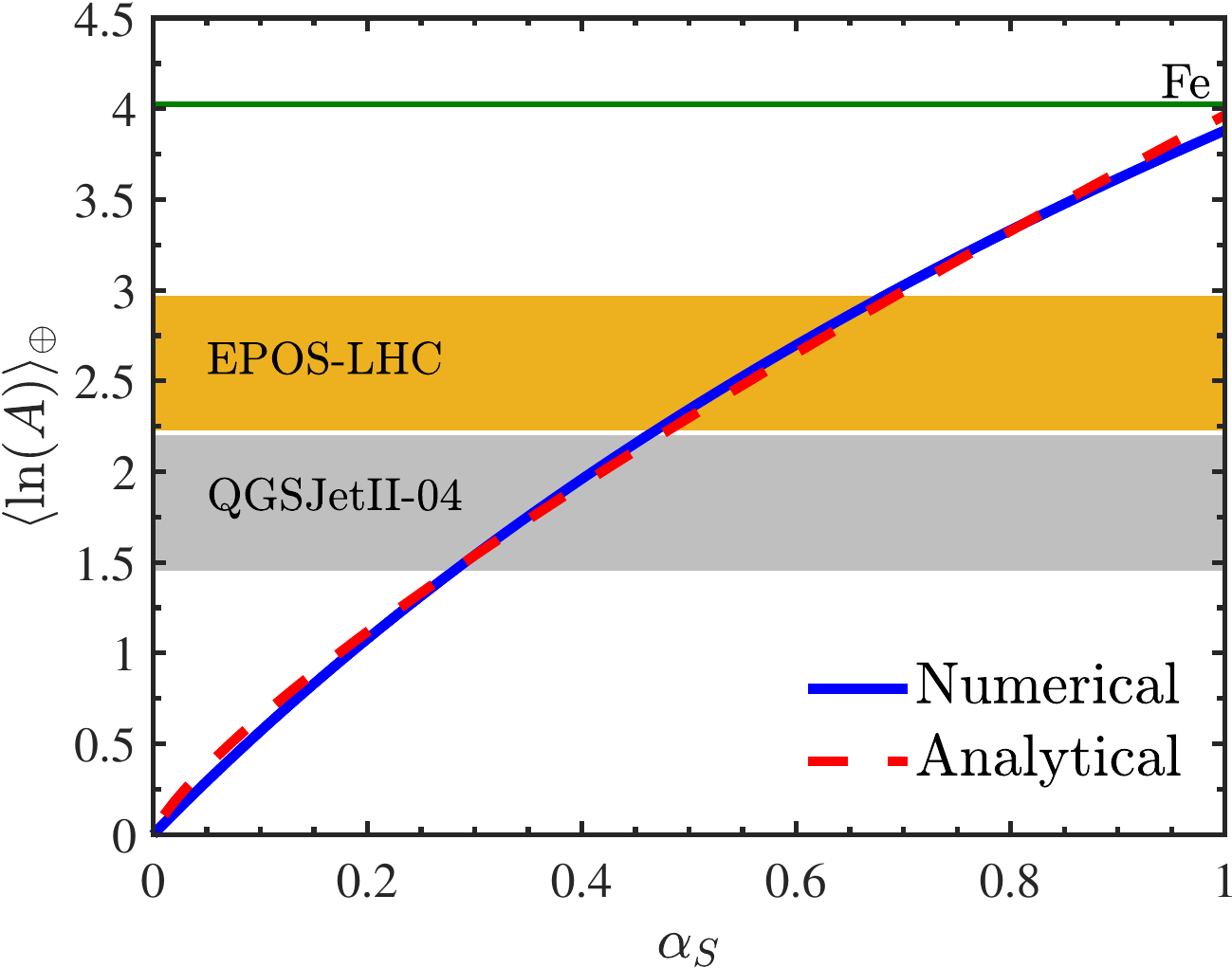}
\caption{Relationship between the composition at the source parameterized by the iron fraction $\alpha_S$ and the mean composition at the Earth $\langle\ln A\rangle_\oplus$.
The red dashed curve shows the fit from Eq.~\ref{eq:alpha2lnA} and the blue curved the correspondent numerical result obtained from CRPropa3. The green line indicates the iron composition for comparison, the yellow and gray bands refer to the Auger composition constraints for $\ECR>32$ EeV obtained by adopting two different hadronic models as from Refs.~\cite{Ostapchenko:2010vb,Pierog:2013ria} to provide an estimate of the hadronic model uncertainty.
}
\label{fig:alpha2lnA}
\end{figure}

Of the four model parameters listed above, we constrain the normalization $\Phi_0$ and the source spectral index $\gamma$ by using the Auger measurement \cite{Fenu:2017hlc} with $\ECR>32$ EeV with bin widths $\Delta\log_{10}\ECR=0.1$ up to $\ECR=160$ EeV. We look for the minimum of the $\chi^2$ test statistic; in all cases, the goodness-of-fit measure is consistent with the Auger data at $<2$ $\sigma$ and we find $\gamma\in[2.5,3]$ for $\alpha_S \in [0,1]$. Since we adopt a flux normalization (the product of the source density and the luminosity) based on the EECR data which comes from $z\lesssim0.02$, $\Phi_0$ is largely independent of $m$.

The two parameters of interest, $\alpha_S$ and $m$, each have some measurements although neither has a robust one. In this work, we leave them free to vary. 

The variable $\alpha_S$ could be usefully parametrized in terms of the composition at the Earth, $\langle\ln A\rangle_\oplus$, which is often adopted by Auger. We find that the two quantities are indeed related by a simple relationship:
\begin{equation}
\langle\ln A\rangle_\oplus=4.0(\alpha_S)^{0.79}\,;
\label{eq:alpha2lnA}
\end{equation}
this is shown in Fig.~\ref{fig:alpha2lnA} where our fit (dashed red curve) is plotted in comparison with the actual curve obtained from CRPropa3 (blue curve).
The currently planned GRAND configuration should be able, in principle, to measure the depth of the shower maximum ($X_{\max}$) of EECRs as well \cite{Fang:2017mhl}.

Converting any measurement of $X_{\max}$ into the actual composition at the Earth relies on hadronic models which are still rather uncertain.
To quantify the size of this uncertainty we also include in the uncertainty estimate the difference between the two main hadronic models used by Auger, QGSJETII-04 \cite{Ostapchenko:2010vb} and EPOS-LHC \cite{Pierog:2013ria}.
With Eq.~\ref{eq:alpha2lnA}, we can relate Auger's composition constraints of $\Delta\langle\ln A\rangle_\oplus=0.75(0.25)$ \cite{Aab:2017cgk} into a constraint on the source composition of $\Delta\alpha_S=0.2(0.1)$ where the hadronic model uncertainty is (is not) included.
Here $\Delta x$ means that $x$ has an uncertainty at one sigma of $\pm\Delta x$. Auger data are then suggesting a composition of the EECR sources which is mixed.

As for the redshift evolution slope, uncertainties on $m$ are currently hard to define given the large systematic errors, see e.g.~Ref.~\cite{Madau:2014bja} and references therein.
Assuming that dedicated statistical fits will be employed in the future in the light of higher statistics, we take an uncertainty on $m$ as $10\%$ at the typical value of $m=3$ for $\Delta m=0.3$, motivated by Ref.~\cite{Strolger:2015kra}.
We will, in general, leave $m$ unconstrained, but in the event that the source population is determined through other measurements or via theoretical arguments, we will adopt a characteristic uncertainty on $m$ of $\Delta m=0.3$.

\section{Results}
\label{sec:results}
In this Section, we present our results on the forecasted event rate in GRAND built with 200,000 radio antennas for different values of the model parameters (i.e., redshift evolution models and EECR source composition). We also discuss the potential of GRAND of discriminating among the different model parameters.

\subsection{Expected cosmogenic neutrino event rate in GRAND}
\begin{figure}
\centering
\includegraphics[width=0.9\textwidth]{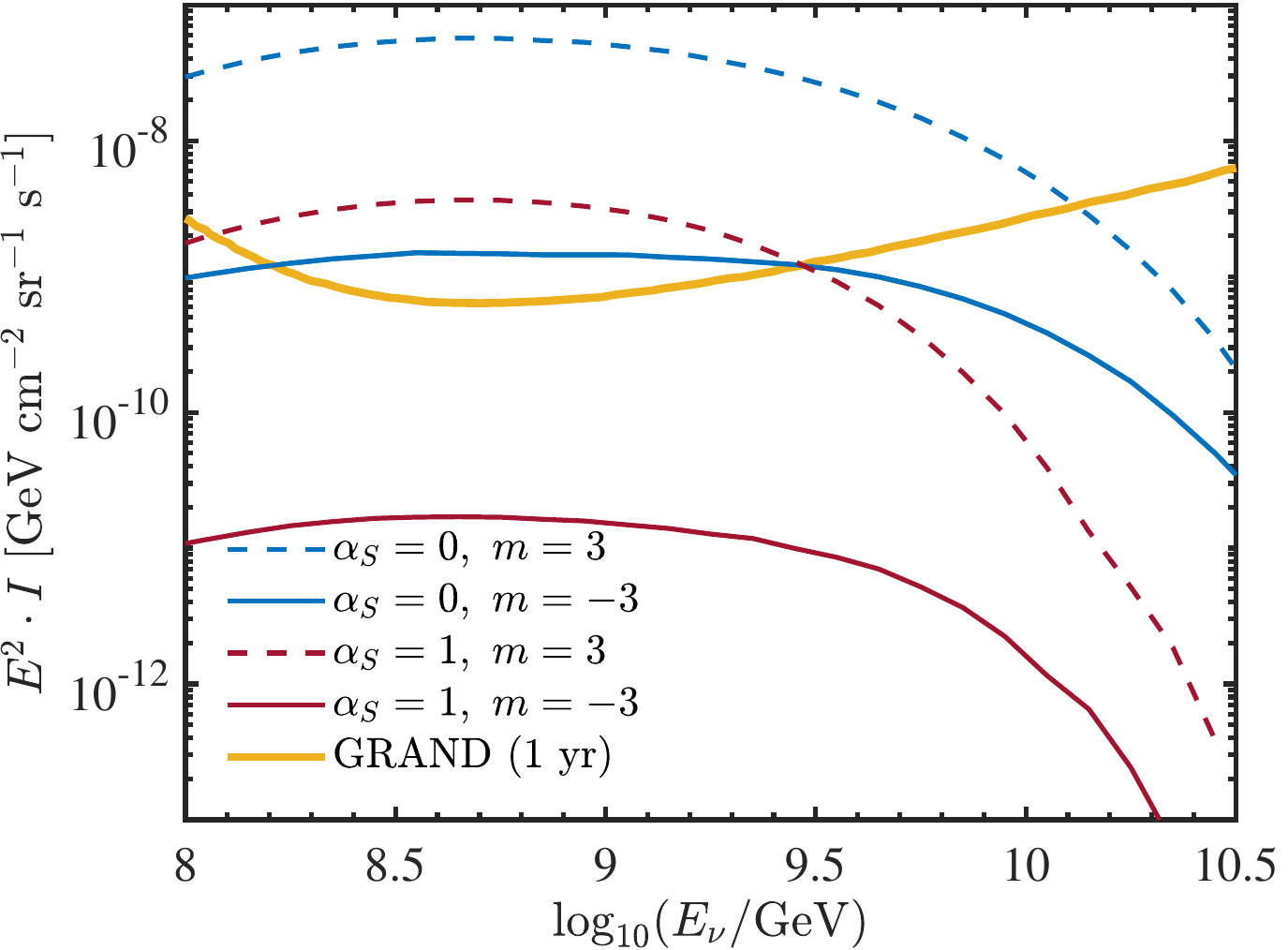}
\caption{Diffuse neutrino intensities for both a proton primary ($\alpha_S=0$, blue curves) and pure iron primary ($\alpha_S=1$, red curves) as a function of the neutrino energy along with the expected sensitivity from GRAND with 200,000 antennas after one year of data taking (orange curve).
The effect of redshift evolution is illustrated through different line types ($m=-3$ with continuous line and $m=3$ with a dashed line). The neutrino intensity is larger as $\alpha$ decreases. The expected intensity of cosmogenic neutrinos is within reach for the currently planned GRAND configuration within the model uncertainties.}
\label{fig:diffuse}
\end{figure}

Figure \ref{fig:diffuse} shows the expected diffuse neutrino intensity for $\alpha_S =0$ (i.e., proton source) in blue and $\alpha_S =1$ (i.e., iron source) in red as a function of the neutrino energy. The dashed (continuous) curves refer to a positive (negative) redshift evolution. 
We see that heavier compositions result in a lower intensity, while stronger redshift evolution leads to higher intensities as expected.
See Appendix \ref{sec:intermediate composition} for a discussion on the composition effect.

The orange curve represents the foreseen sensitivity for GRAND if built with 200,000 antennas from Ref.~\cite{GRAND_whitepaper}. One can see that the expected intensity of cosmogenic neutrinos will be within reach for GRAND for all but the most pessimistic of cases (heavy composition and negative source evolution).

\begin{figure}
\centering
\includegraphics[width=\textwidth]{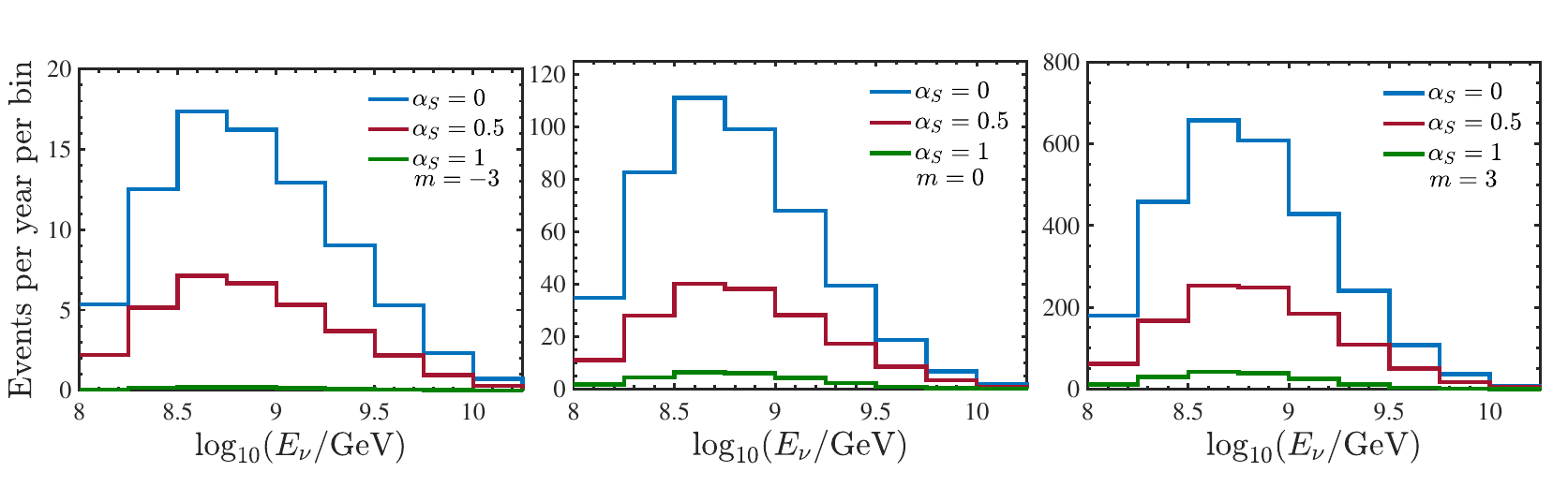}
\caption{Expected event rates seen at GRAND, if designed as foreseen in Ref.~\cite{Martineau-Huynh:2015hae}, per year as a function of the neutrino energy.
The left, middle, and right panels correspond to negative, no, and positive redshift evolution respectively ($m=-3,0,3$).
The blue, red, and green curves correspond to proton, mixed, and iron source compositions respectively ($\alpha_S=0,0.5,1$). The expected event rate is larger as $m$ increases and $\alpha$ decreases.
}
\label{fig:rates}
\end{figure}

Using GRAND's effective area \cite{GRAND_whitepaper} we can convert the diffuse intensities into event rates in order to perform a sensitivity statistical analysis.
These are shown in Fig.~\ref{fig:rates}. The plot on the left corresponds to the negative redshift evolution scenario ($m=-3$), the plot in the middle to the no evolution case ($m=0$) and the one on the right to the case of positive redshift evolution ($m=3$). For each panel, the three curves represent the event rates in the case of proton, mixed, and iron source composition cases respectively (i.e., $\alpha_S=0,0.5,1$). Similarly to what shown in Fig.~\ref{fig:diffuse}, one can see that the expected event rate is larger as $m$ increases and $\alpha$ decreases.

\subsection{Constraints on the source redshift evolution and composition}
\begin{figure}
\centering
\includegraphics[width=.9\textwidth]{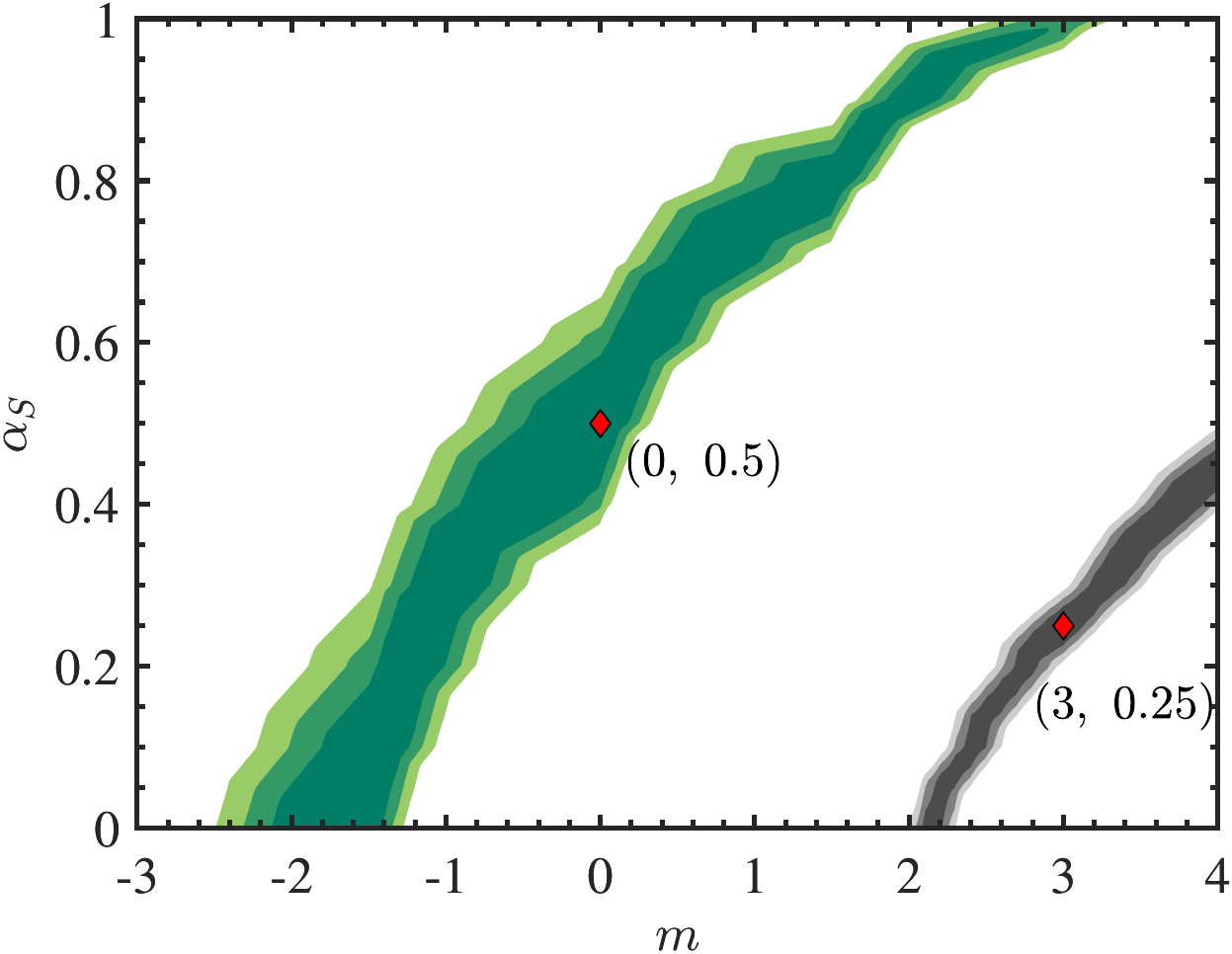}
\caption{The 1,2,3 $\sigma$ allowed region for 2 d.o.f.~given true values of $(m,\alpha_S)=(0,0.5)$ in green and $(m,\alpha_S)=(3,0.25)$ in gray.
This shows that $\alpha_S$ and $m$ are fully degenerate and neither can be determined from a cosmogenic neutrino measurement alone.
}
\label{fig:2D}
\end{figure}

Next we calculate the 1,2,3 $\sigma$ allowed regions for various assumptions on the source composition parameter $\alpha_S$ and redshift evolution slope $m$. Assuming as true values $(m,\alpha_S)=(0,0.5)$ and $(m,\alpha_S)=(3,0.25)$, Fig.~\ref{fig:2D} shows contour plots of the 1,2,3 $\sigma$ allowed regions for 2 d.o.f.~in the $(m,\alpha_S)$ plane obtained by assuming the currently planned GRAND design \cite{Martineau-Huynh:2015hae}. One can see that $\alpha_S$ and $m$ are fully degenerate with one another.
This degeneracy is valid for any $(m,\alpha_S)$ pair as evident from the green and gray bands, although stronger redshift evolutions and lighter compositions result in narrower bands as they lead to more statistics.

By increasing the amount of input information, the degeneracy between the redshift evolution slope and composition can be broken.
For example, if we assume that the redshift evolution slope is known with an uncertainty of $\Delta m=0.3$, we can strongly constrain the composition to $\Delta\alpha_S\in[0.02,0.18]$.
Alternatively, if we know the source composition parameter from $X_{\max}$ or other measurements at the level of $\Delta\alpha_S=0.2(0.1)$ then the redshift evolution slope can be constrained to $\Delta m\in[0.2,1.3]$ ($[0.1,1.2]$) where the first estimate assumes that the hadronic uncertainties remain the same and the second that they are significantly reduced in the coming years.
These results are summarized in Table \ref{tab:1D results}. Notably, given the higher event statistics (see Fig.~\ref{fig:rates}), we find that the error in the determination of $\alpha_S$ is smaller as $m$ increases, while the uncertainty in the measurement of $m$ increases as $\alpha_S$ increases. 

\begin{table}
\centering
\caption{The one sigma uncertainties on $\alpha_S$ (left) and $m$ (right) with external inputs and for various true values.
The uncertainty on $\alpha_S$ in the left table assumes $m$ is known to $\Delta m=0.3$.
The uncertainty on $m$ in the right table assumes $\alpha_S$ is known to $\Delta\alpha_S=0.1$, that is, that the hadronic uncertainties are reduced to be below the experimental uncertainties on $X_{\max}$. The uncertainty of $\alpha_S$ ($m$) is smaller (larger) as $m$ ($\alpha_S$) increases.}
\label{tab:1D results}
\begin{tabular}{c|c|ccc}
\multicolumn{2}{c|}{\multirow{2}{*}{$\Delta\alpha_S$}}&\multicolumn{3}{|c}{$\alpha_S$}\\\cline{3-5}
\multicolumn{2}{c|}{}&0&0.5&1\\\hline
\multirow{3}{*}{$m$}&-3&0.18&0.09&0.03\\
&0&0.15&0.08&0.02\\
&3&0.06&0.05&0.02\\
\end{tabular}
\qquad\qquad\qquad
\begin{tabular}{c|c|ccc}
\multicolumn{2}{c|}{\multirow{2}{*}{$\Delta m$}}&\multicolumn{2}{|c}{$\alpha_S$}\\\cline{3-5}
\multicolumn{2}{c|}{}&0&0.5&1\\\hline
\multirow{3}{*}{$m$}&-3&0.19&0.34&0.38\\
&0&0.15&0.41&1.22\\
&3&0.19&0.34&0.24\\
\end{tabular}
\end{table}

We also examined the effect of improving the energy resolution to extremely optimistic experimental parameters of $\Delta\log_{10}E_\nu=0.1$ which is equivalent to $^{+26\%}_{-21\%}$.
This resulted in no improvement in the constraints due to the fact that the shape differences between the different cases are much too small to be probed. Thus, interestingly, the ability of GRAND to constrain EECR source parameters comes almost entirely from the event rate.

\section{Discussion} \label{sec:discussion}
Our results differ than others in the literature, see e.g.~Ref.~\cite{AlvesBatista:2018zui}, due to the fact that we fit the spectral index to the Auger data which results in a peak neutrino event rate at roughly $\log_{10}(E_\nu/$GeV$)\in[8.5-8.75]$, see Fig.~\ref{fig:rates}.
The location of this peak is \emph{largely independent} of the source composition and redshift evolution slope which explains why these parameters are fully degenerate with each other, even with exceptional energy resolution.
The composition has a relatively small shape effect for two reasons.
The first is because fitting the source spectral index accommodates some of the difference induced by varying the composition.
The second reason is because GRAND, as currently designed, will be sensitive in a relatively modest energy region (see e.g.~Fig.~\ref{fig:rates}).

One caveat is that while our analysis is independent of the large hadronic model uncertainties prevalent in $X_{\max}$ and $X_{\max}^\mu$ (the depth of the air shower determined from the muonic component) measurements, there are also uncertainties due to the EBL and the photodisintegration models \cite{Batista:2015mea,Boncioli:2016lkt}.
Our understanding of the EBL continues to become more and more precise and photodisintegration models may also improve in the future. Notably, the EBL model adopted here~\cite{Finke:2009xi} does not differ much from the one recently presented in Refs.~\cite{Ajello:TeVPA18,Ajello:2018sxm}, suggesting that our conclusions are robust and that these uncertainties may be under control in the future.

Another relevant factor affecting the composition is the metallicity of the host environment. In fact it may well be that composition correlates with metallicity which likely scales with redshift \cite{2018MNRAS.473.3312A,2017arXiv171105261T}.
We have assumed in this article that the composition evolution is flat since the uncertainties on metallicity dependence are large and because it is unclear how metallicity maps onto EECR composition.
Moreover, including such a dependence would provide an additional parameter that would also be degenerate with the others indicating that extracting information on metallicity evolution from the cosmogenic neutrino flux would be extremely difficult without combining several independent data sets.

As for the redshift evolution of the EECRs sources, very strongly evolving sources such as Active Galactic Nuclei (AGNs) with $m\sim5$ \cite{Stanev:2008un} are already somewhat in tension with Auger's constraints on the ultra high energy neutrino flux \cite{Aloisio:2015ega}.
This justifies our focusing on $m<4$.
In any case, if GRAND will be built with 200,000 antennas, it would easily be able to make a measurement with excellent precision if the flux was just under the current limits due to a very strong redshift evolution and would provide enough information to identify such a strong redshift evolution.

It will be crucial to know if the ultra high neutrinos measured by GRAND will be coming from EECRs or from an extension of the non-thermal flux already measured by IceCube \cite{Aartsen:2018vtx}.
Determining the exact origin of the neutrinos eventually detected from GRAND requires good enough energy resolution, $\Delta\log_{10}E_\nu\lesssim0.25$, especially in the lower energy tail of GRAND sensitivity. 
Other experiments such as POEMMA which is sensitive at lower energies than GRAND may provide a key component to connecting the flux measured by GRAND to that from IceCube and may further help in determining if the flux measured by GRAND is cosmogenic or not.
In the event that there is a new flux of ultra high energy neutrinos separate from that measured by IceCube and right within the cosmogenic window, however, identifying the cosmogenic component will be extremely difficult from spectral measurements alone; point source identification may be the only way forward in this case.

Other experimental programs such as ARA \cite{Allison:2015eky} and ARIANNA \cite{Nelles:2016fxe} are sensitive to the cosmogenic neutrino flux, but only in the most optimistic scenarios (i.e., very strong evolution $m\gtrsim3$ and light composition) which seem to be disfavored by Auger data.
Although their sensitivity to estimating EECR parameters will be lower than the one currently foreseen for GRAND since the event rates will be $<1$ for the majority of the relevant parameter space, a combined analysis of all data could provide robust hints on the global source properties studied in this work.

\section{Conclusions}
\label{sec:conclusions}
The sources of extreme cosmic rays with energies above $55$ EeV are not yet known. A guaranteed flux of cosmogenic neutrinos is expected in association to the extreme energy cosmic ray flux. Those neutrinos will be measured by next generation radio facilities such as the Giant Radio Array for Neutrino Detection (GRAND) and the Probe Of Extreme Multi-Messenger Astrophysics (POEMMA).

In an effort to understand what information we will learn about the sources accelerating cosmic rays to extreme energies, we have fixed the overall normalization of the cosmogenic neutrino intensity as well as the slope of the injection spectral index by fitting the cosmic ray spectrum to the Auger data~\cite{Fenu:2017hlc} and calculated the sensitivity of GRAND, if built with 200,000 radio antennas, to global source parameters such as the redshift evolution law and the average source composition. Our findings suggest that there is no significant shape difference in the expected neutrino intensities among different compositions and redshift evolution slopes.
Hence, GRAND, even in its most optimistic design, cannot break the degeneracy between the redshift evolution slope and the composition, regardless of the energy resolution.

Several assumptions were made in this analysis.
The rigidity ansatz was made to limit the number of free parameters in the astrophysical fit, as is not expected to significantly alter our results.
In addition, we fix the source redshift evolution to a parametric form where only the $z<1$ part is allowed to vary.
This generally reproduces a variety of different redshift evolutions commonly used, and any variation in the high redshift part does not significantly affect our results.
Next, we parameterize the composition as a function of $p$ and Fe content only.
In principle a multi-composition model can modify the resultant fit, but most cases should be somewhere within the scope of our parameterization.
The assumption that has the largest effect on our results is that there is no significant additional ultra high energy neutrino flux in this energy region.
If another class of sources with a comparable or larger flux of neutrinos is present, then constraining the composition and redshift evolution will become extremely difficult.

Interestingly, by assuming a prior on the variability range of one or the other parameter, this degeneracy can be broken. By assuming only one year of data taking of GRAND with 200,000 antennas, the composition of the sources of ultra high energy cosmic rays would be determined at a $\sim5-10\%$ level given that the redshift evolution is known by other means.
For example, a direct identification of cosmic ray sources combined with the current redshift evolution measurements of various source classes.
Similarly, the redshift evolution can be constrained at a $10\%$ level given that the composition is well measured perhaps by $X_{\max}$ measurements from the Pierre Auger Observatory \cite{Fenu:2017hlc}, Telescope Array \cite{Ikeda:2017dci}, and GRAND \cite{Fang:2017mhl}.
Both of these are comparable to or better than optimistic current uncertainties which carry different systematics.
This makes the measurement of the cosmogenic neutrino flux a crucial and very orthogonal measurement to constrain the properties of EECRs and their sources even as future measurements of the redshift evolution of the star formation rate continues to improve from IR and H$\alpha$ measurements \cite{Calzetti:2007md,Martinache:2018zgg,Smit:2015aif}.

Next generation large scale radio facilities and fluorescence satellites, such as GRAND and POEMMA respectively, will be therefore crucial.
They have the potential to not only lead to the detection of the long-hunted flux of cosmogenic neutrinos, but also to help us to unveil the nature of cosmic accelerators through their neutrinos.

\acknowledgments
We thank Rafael Alves Batista and Kumiko Kotera for helpful comments.
PBD and IT acknowledge support from the Villum Foundation (Project No.~13164), and by the Danish National Research Foundation (DNRF91).
PBD thanks the Danish National Research Foundation (Grant No.~1041811001) for support.
The work of IT has also been supported by the Knud H\o jgaard Foundation and the Deutsche Forschungsgemeinschaft through Sonderforschungbereich SFB 1258 ``Neutrinos and Dark Matter in Astro- and Particle Physics'' (NDM).

\appendix

\section{Sources with Intermediate Composition}
\label{sec:intermediate composition}
With regards to the source composition, we have been acting under the assumption that the neutrino flux is monotonically decreasing as the composition increases from proton to iron.
For heavy nuclei this is because heavier nuclei have a smaller amount of energy per nucleon, and thus typically a smaller neutrino energy.

\begin{figure}
\centering
\includegraphics[width=.9\textwidth]{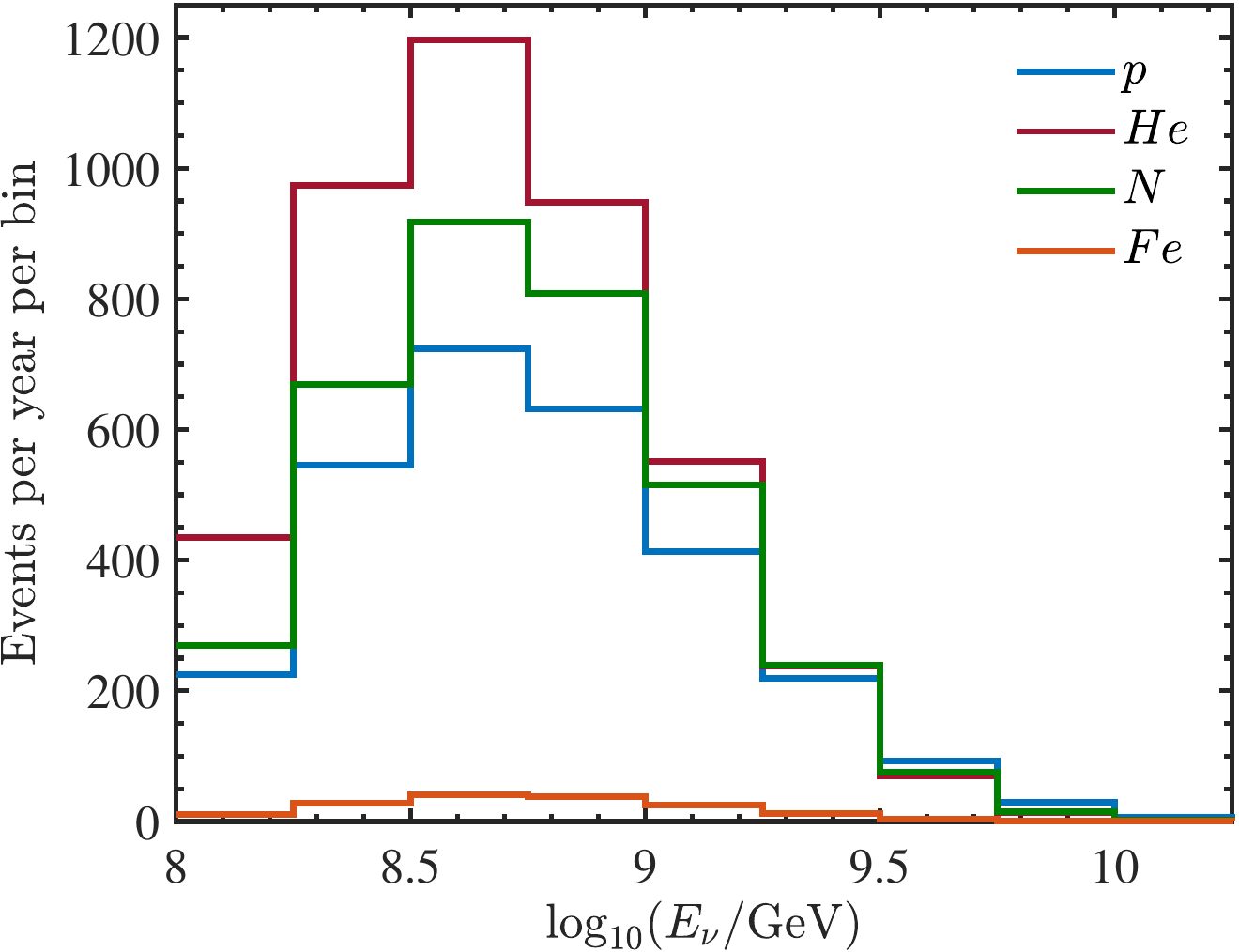}
\caption{Event rates at GRAND for different source compositions as a function of the neutrino energy for $m=3$.
Contrary to conventional expectation, the neutrino flux is larger for light to intermediate nuclei than for protons when the EECR spectrum is required to fit the data.
}
\label{fig:intermediate composition}
\end{figure}

This picture is only approximate within our framework.
In fact, under the assumption that we are required to reproduce the EECR spectra measured by Auger, a pure He or N source composition results in a larger neutrino flux than the one obtained assuming a pure proton composition at the source, although the pure N case does not result in a good fit to the Auger's EECR spectral data.
This is shown in Fig.~\ref{fig:intermediate composition}.

Note that this is somewhat different than some other results in the literature, e.g.~Refs.~\cite{Aab:2016zth,AlvesBatista:2018zui}, since we are not including $X_{\max}$ or $X_{\max}^\mu$ information and are parameterizing the composition with a single number instead of with several free parameters.
This is required to be able to determine what GRAND's composition sensitivity will be.

Despite this fact, we have chosen to parameterize the composition in terms of the proton and iron fraction to agree with the convention commonly adopted in the literature.
In reality, the parameterization of Eq.~\ref{eq:alphaS} is conservative. In fact, if we had used He and Fe, the $\alpha_S$ sensitivity would be better given the foreseen larger event rate.
That is, values of $\alpha_S>1$ are allowed, although then the definition of $\alpha_S$ is no longer derived from Eq.~\ref{eq:alphaS} and is related to Fig.~\ref{fig:intermediate composition}.

\bibliographystyle{JHEP}
\bibliography{UHEnus}

\end{document}